\documentclass{article}

\usepackage{PRIMEarxiv}
\usepackage[utf8]{inputenc}
\usepackage[T1]{fontenc}
\usepackage{hyperref}
\usepackage{url}
\usepackage{booktabs}
\usepackage{amsfonts}
\usepackage{nicefrac}
\usepackage{microtype}
\usepackage{fancyhdr}
\usepackage{graphicx}
\usepackage{dirtytalk}
\usepackage{listings}
\usepackage{longtable}
\usepackage{xcolor}
\usepackage{array}
\usepackage{float}
\usepackage{minted}
\usepackage{placeins}
\usepackage{hyperref}

\usepackage{cite}

\title{\textbf{Benchmarking Large Language Models for ABAP Code Generation: An Empirical Study on Iterative Improvement by Compiler Feedback\\\phantom{x}\\Preprint}}

\author{
  \begin{tabular}{cc}
    \begin{minipage}[t]{0.45\textwidth}
      \centering
      \textbf{Stephan Wallraven} \\
      \small \textmd{Technische Hochschule Köln} \\
      \small \texttt{\href{mailto:stephan_andreas.wallraven@th-koeln.de}{\nolinkurl{stephan_andreas.wallraven@smail.th-koeln.de}}}
    \end{minipage} 
    & 
    \begin{minipage}[t]{0.45\textwidth}
      \centering
      \textbf{Tim Köhne} \\
      \small \textmd{Technische Hochschule Köln} \\
      \small \texttt{\href{mailto:tim.koehne@smail.th-koeln.de}{\nolinkurl{tim.koehne@smail.th-koeln.de}}}
    \end{minipage} 
    \\[1cm]
    \begin{minipage}[t]{0.45\textwidth}
      \centering
      \textbf{Hartmut Westenberger} \\
      \small \textmd{Technische Hochschule Köln} \\
      \small \texttt{\href{mailto:hartmut.westenberger@th-koeln.de}{\nolinkurl{hartmut.westenberger@th-koeln.de}}}
    \end{minipage} 
    & 
    \begin{minipage}[t]{0.45\textwidth}
      \centering
      \textbf{Andreas Moser} \\
      \small \textmd{CONSILIO GmbH} \\
      \small \texttt{\href{mailto:andreas.moser@consilio-gmbh.de}{\nolinkurl{andreas.moser@consilio-gmbh.de}}}
    \end{minipage}
  \end{tabular}
}

\begin{document}
\maketitle

\begin{abstract}
This work investigates the performance of Large Language Models (LLMs) in generating ABAP code. Despite successful applications of generative AI in many programming languages, there are hardly any systematic analyses of ABAP code generation to date. The aim of the study is to empirically analyze to what extent various LLMs can generate syntactically correct and functional ABAP code, how effectively they use compiler feedback for iterative improvement, and which task types pose special challenges. For this purpose, a benchmark with 180 tasks is conducted, consisting of adapted HumanEval tasks and practical SAP scenarios. The results show significant performance differences between the models: more powerful LLMs achieve success rates of around 75\% after several iterations and benefit greatly from compiler feedback, while smaller models perform significantly weaker. Overall, the study highlights the high potential of powerful LLMs for ABAP development processes, especially in iterative error correction.
\end{abstract}


\section{Introduction}
\subsection{Background and Motivation}
Over the course of the rapid further development of LLMs in recent years, code generation is increasingly finding its way into the software development process. ChatGPT, Claude-Sonnet, or specific software development tools like GitHub Copilot impressively show that generative AI systems are capable of automatically generating, explaining, and optimizing program code \cite{e25060888}.\label{sec:TraingsdatenPython} A major focus of these developments has so far been on widely used programming languages such as Python or JavaScript, for which large amounts of training data and benchmarks are available. In these languages, LLMs have already demonstrated a high level of correctness in the generated programs \cite{10.1145/3597503.3623316, chen_evaluating_2021, cao_javabench_2024}. However, domain-specific enterprise languages such as ABAP, which plays a central role in SAP systems, have received much less research attention so far. The use of AI for automated code generation promises significant economic benefits, as development processes can be designed more efficiently and routine tasks can be substantially reduced.

SAP is one of the leading providers in the ERP market and significantly shapes the digital transformation of companies. ERP systems like SAP are essential for controlling business processes and integrating business logic in companies of all sizes. For many SAP systems, ABAP remains the central programming language for adapting and extending business logic. The high prevalence of SAP systems and the large ABAP community make any productivity increase, such as through AI-supported code generation, relevant and economically significant for many companies \cite{SAPUmsatzWeltweit}.

\subsection{Problem Statement}
Despite general progress in the field of generative AI, the application of LLMs for ABAP code generation has hardly been researched thus far. As a domain-specific and proprietary language, ABAP poses a particular challenge because it is deeply embedded in the architecture and semantics of the SAP ecosystem. It features specific language concepts, such as internal tables, data declarations, modularized program structures, and close interaction with SAP transactions and databases, which might be insufficiently grasped by general LLMs due to a potential lack of domain knowledge.

This insufficient domain knowledge presumably represents a central problem, which can be attributed, among other things, to the lack of availability of suitable training data. For example, Peter Färbinger writes in a blog post from 2024 \cite{faerbingerSAPKIDystopieE3Magazin2024}: \say{The more training data an AI receives, the more intelligent it can become. An example: SAP is currently still failing with a Co-Pilot for ABAP, SAP's own programming language. Worldwide, there is far too little ABAP program code to train a generative AI like ChatGPT from OpenAI.} The statement that ABAP is a \say{low-resource language} due to a lack of public training data and that generative AI models therefore often perform worse than with languages like Python or Java is confirmed in current blog posts by SAP and their partners. A working group of SAP and NVIDIA employees notes in a blog post from 2025: \say{Because ABAP is a closed-source and low-resource language, general purpose commercial LLMs often struggle to produce correct syntax and follow ABAP cloud programming best practices. Their performance is subpar compared to their capabilities in more commonly used languages such as Python and Java.} \cite{sohrabij_sap_2025}.

In addition to data availability, there are methodological challenges in evaluating generated ABAP programs. Since the language is strongly typed and relies on the SAP ABAP runtime environment, the automatic assessment of syntactic and semantic correctness is complex. Furthermore, it remains unclear to what extent LLMs can actually generate functionally correct and practically relevant ABAP code, especially in scenarios with multiple tables, transactions, or logical dependencies. Despite the enormous economic importance, AI-supported ABAP code generation is still in its infancy. For example, one of the few existing studies on code generation in ABAP shows limited results, which are often based on few test cases or purely manual evaluations (see Chapter \ref{sec:related_work}). 
This results in a research gap regarding the systematic and empirical investigation of the performance of LLMs in generating syntactically and semantically correct ABAP code. Previous works do not provide reliable statements on reliability, error rates, or the limits of generative models in this domain-specific environment.

\subsection{Objectives and Research Questions}

This work aims to apply a benchmark approach to ABAP code generation as a prerequisite to empirically investigate and systematically evaluate the performance of LLMs in generating ABAP code. By such an environment it can be examined to what extent LLMs can produce ABAP programs that are syntactically correct, semantically accurate, and practically relevant code. The focus lies on a comparative analysis of different models with respect to their capability to correctly apply simple language constructs, such as internal tables, data declarations, and modularization concepts which are critical elements in real SAP development scenarios. Generated programs will be assessed using defined evaluation metrics, including correctness and functionality. To enable reproducible evaluation, a standardized test environment will be developed. Furthermore, an analysis of errors in the generated code will be conducted to identify typical weaknesses and systematic patterns in code generation. Another objective is to explore whether code generation can be improved through an iterative process that incorporates compiler error messages as feedback.  

Drawing from these objectives, the following research questions guide the study:  
\begin{enumerate}
    \item To what extent can LLMs generate syntactically and semantically correct ABAP code?  
    \item Can incorporating ABAP compiler error messages improve code quality? 
    \item Which programming tasks present particular challenges for ABAP code generation?
\end{enumerate}

\section{Related Work}
\label{sec:related_work}
The application of LLMs in the field of automated code generation has become increasingly relevant in recent years. Various studies suggest that LLMs may be able to generate code directly from descriptions, potentially allowing even less experienced developers to create functional programs by formulating requirements in natural language \cite{10.1145/3747588, 10.1145/3597503.3623316}.

HumanEval developed by Chen et al. has established itself as one of the central benchmarking procedures for the systematic evaluation of the code generation capabilities of such models. Due to its clear structure and broad reception, the framework is now considered a standard metric for program synthesis and often forms the basis for comparing different model architectures \cite{wang2025softwaredevelopmentlifecycle}. It consists of programming tasks formulated in natural language and validated by predefined unit tests, thus enabling an objective measurement of functional correctness \cite{chen_evaluating_2021}.

While HumanEval primarily evaluates the general capabilities of LLMs for code generation, the question arises for this work in particular how such approaches can be used in specific corporate environments such as SAP. In the SAP context, tools such as the AWS SAP ABAP Assistant or SAP Joule for Developers are intended to support the generation of ABAP code from natural language prompts and thus extend the possible applicability of LLMs for practical development tasks \cite{figueiredo_ai_2025}.

The study by Gałązka and Szymański on ABAP code generation that LLMs can generate not only simple programs but also more complex ABAP objects using modern language features \cite{soliman_using_2025}. However, a large part of the previous investigations is based on limited datasets, so that the transferability of the results to more extensive or productive scenarios can currently only be evaluated to a limited extent. Studies also indicate that by integrating LLM models into SAP systems, certain development tasks, such as generating code artifacts or tests, can be partially automated, while chatbot-like forms of interaction facilitate the processing of individual, context-specific tasks \cite{figueiredo_ai_2025, soliman_using_2025}.

A frequently discussed advantage of such approaches lies in the potential increase in efficiency. According to current findings, generative AI can accelerate the transition from conceptual descriptions to executable code and automate certain routine tasks, such as report generation \cite{sohrabij_sap_2025, soliman_using_2025}. In addition, it is reported that tools like the AWS SAP Assistant or Joule for Developers can support the documentation and analysis of legacy code, which could contribute to reducing project runtimes \cite{figueiredo_ai_2025}. Various technical procedures are used to ensure the quality of the generated programs. Research results on constrained decoding indicate that token-based validation using ABAP grammar can help generate syntactically correct suggestions and increase success rates for cloud tasks \cite{sohrabij_sap_2025}. In addition, the importance of well-structured prompts is emphasized in order to obtain context-sensitive and consistent results. The quality of the training data and the chosen model configuration are also considered central influencing factors on the performance of the models \cite{soliman_using_2025}.

Despite these developments, there are still significant limitations. Studies show that LLMs show performance limits especially in complex, novel, or cross-system programming tasks and that human expertise therefore remains necessary \cite{10.1145/3747588, soliman_using_2025}. In addition, security-relevant and ethical questions in connection with automatically generated code are discussed \cite{10.1145/3747588}. Procedures such as Nvidia Inference Microservices can optimize latency and throughput, but at the same time make it clear that validation and performance optimization remain central challenges for the productive use of LLM-based code generation \cite{sohrabij_sap_2025}.

\section{Methodology}
\begin{figure}[h!]
    \centering
        \includegraphics[width=0.65\textwidth]{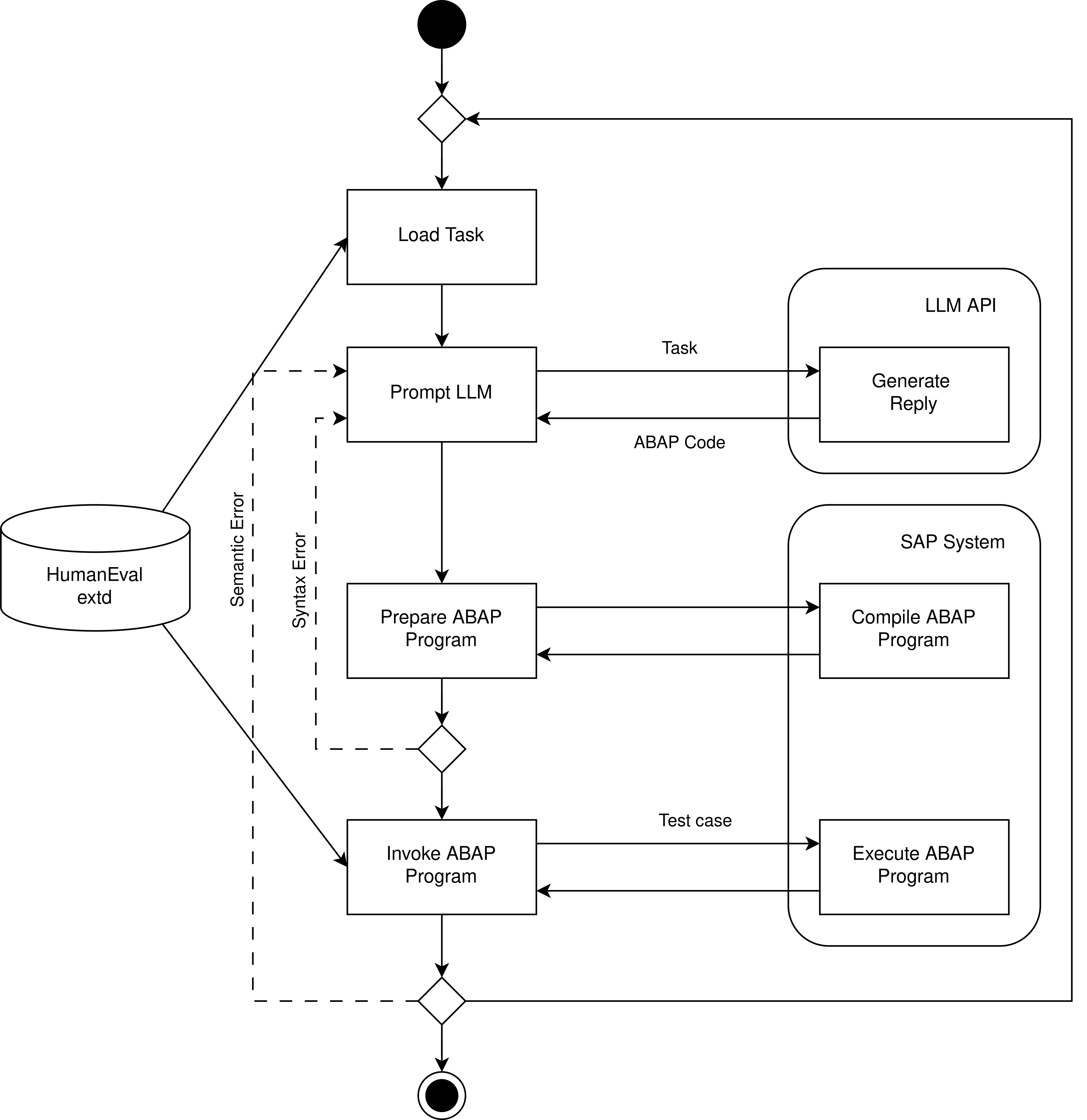}
        \caption{Rough Process of the Benchmark.}
        \label{fig:uebersicht}
\end{figure}

The experimental procedure for the empirical investigation of ABAP code generation by LLMs is based on several central aspects:  
A benchmark with 180 tasks serves as the data basis, covering both general algorithmic problems (based on HumanEval) and specific SAP scenarios. The tasks are divided into five thematic categories, including for example \textit{String Handling} and \textit{ABAP Database Operation}. A special focus is placed on increasing the probability of successful code generation through an iterative feedback process: each task is processed in up to five feedback loops, during which the system provides the LLM with feedback from the ABAP compiler or automated unit tests. Success is measured based on functional correctness, which is verified by these automated unit tests. In addition, a detailed error analysis and a Kaplan-Meier survival analysis are carried out to evaluate the learning curves. The entire process is fully automated and uses a standardized SAP environment (Docker image), which is addressed via a Python control and the ADT interface. This ensures the reproducibility of the results.

Figure \ref{fig:uebersicht} graphically shows the basic sequence of the benchmark. The structure and methodology of the benchmark are described in detail below.

\subsection{Structure and Methodology of the Benchmark}
\label{sec:methodik_benchmark}

The developed benchmark comprises a total of 180 test tasks, which are divided into two groups: 164 standardized tasks based on the HumanEval dataset \cite{chen_evaluating_2021}, which was originally developed for evaluating code generation in Python, and 16 ABAP-specific tasks. These 16 specific tasks are derived from practical SAP-specific scenarios focusing on the processing of database tables. In addition, five task categories are created  and each task is assigned by hand to one or more categories. The category \textit{String Handling} includes 76 tasks, \textit{List or Array Operation} 108 tasks, \textit{Mathematical Calculation} 93 tasks, \textit{Logical Condition} 125 tasks, and \textit{ABAP Database Operation} 16 tasks. The dataset for the benchmark is available in the appendix \ref{huggingface}.

This categorization allows for a differentiated analysis of model performance in relation to different problem types and reflects both algorithmic foundations and real ABAP use cases.
The test cases cover different levels of difficulty and types of tasks, enabling a broad evaluation of the models.\label{sec:systemprompt_erklaerung} 
A constant system prompt (see code example \ref{lst:abap-systemprompt}) serves as the basis for interaction in order to clearly structure the interaction, unambiguously specify the task, and promote the consistency of the generated answers. The specification of the system prompt is specifically geared towards ABAP to avoid deviations in implementation. The following aspects were considered:
Global classes with static methods are required, as the test setup requires this structure and this ensures the comparability of the results.
The definition of a uniform version of the SAP runtime environment prevents version differences from leading to misinterpretations or incorrect code generations. The standardized decimal separator ensures consistent processing of numeric values, as ABAP theoretically supports both variants. 
Development rules such as RETURNING parameters, the restriction to one public method per class, and naming conventions (classes start with \say{Z}) were implemented to ensure uniformity, traceability, and automatability of the tests. Both class definition and implementation code are to be provided, as both are required for carrying out the evaluation. 
Exclusively returning the code without additional explanations reduces automation-related errors and enables easier processing in the test system. These targeted specifications ensure that the test execution is reproducible, valid, and systematically comparable. Each task is repeated a total of ten times to allow for statistically robust evaluations given the stochastic nature of LLMs. Up to five feedback loops are provided in each individual task solution process. They serve to give the LLM the opportunity to improve errors, and the number five is based on a realistic estimate of inquiries that a developer would also make manually.

The benchmark is carried out on several LLMs. The inputs are limited in length depending on the model: For models without reasoning functionality, the maximum response length is set to 5000 tokens per message, while for reasoning models it is increased to 10000 tokens, as these models typically generate more extensive outputs. The output of the models is limited to avoid infinite or extremely long answers, but at the same time leave enough room for complete solutions. The temperature parameter is set to a value of 0.2 for all models (with the exception of GPT-5) to ensure increased predictability and consistency of the generated answers. This is particularly important in the context of code generation, as stable and deterministic outputs facilitate verifiability and reproducibility of results. However, the influence of the temperature parameter on model performance and the quality of the generated outputs represents a relevant aspect that should be systematically analyzed in further investigations. For GPT-5 the temperature cannot be changed due to restrictions by OpenAI so we could not use the same setting.

\subsection{Selection of LLMs}
A selection of different LLMs is made for carrying out the benchmark, covering different properties and focuses. The choice of LLMs does not aim to comprehensively test all available models but rather focuses on a comparative analysis between selected open-source models and leading closed-source models that are among the most widely used LLMs in current research. The goal is to investigate fundamental scaling effects and architectural differences in the context of ABAP code generation.

A focus is on model diversity: Both open-source models like Llama-3.3-70B-Instruct and Codestral-22B, as well as closed-source models like GPT-5 and Claude-Sonnet-4 were included to enable comparison between freely accessible and proprietary approaches. In addition, attention was paid to diversity in architecture and model size: The range extends from smaller models with around 7 billion parameters (like Qwen2.5-Coder) to large open models like GPT-OSS-120B. This is complemented by proprietary models like GPT-5 and Claude-Sonnet-4, whose parameter numbers are estimated to be significantly higher and which cover further capacity levels. This allows conclusions to be drawn about possible scaling effects and their influence on the quality of the generated answers.

Another aspect is the degree of specialization of the models. Both general-purpose LLMs like GPT-5, which were trained for diverse tasks, and code-specifically trained models like Qwen3-Coder are considered. This combination enables comparison between universally applicable models and those specifically optimized for code generation. Furthermore, all models used are available as instruct variants, which is particularly relevant for testing code generation under realistic prompting conditions. The majority of the selected models also belong to the most well-known LLMs frequently used in research, which facilitates comparability with other scientific investigations. By using established models, the results can be classified and verified in relation to existing studies, increasing the validity and comprehensibility of the analysis.

Ultimately, the selection falls on the following models: Llama-3.3-70B-Instruct, Codestral-22B, Qwen2.5-Coder and Qwen3-Coder, GPT-OSS-20b and GPT-OSS-120b, aswell as GPT-5-2025-08-07 and Claude-Sonnet-4-20250514. The benchmark evaluations were conducted in August and September 2025.

\subsection{Evaluation Environment}
To ensure reproducibility, the official SAP Docker image \mbox{\textit{sapse/abap-cloud-developer-trial:2023}} is used as the basis, which provides a standardized ABAP development environment. This gives us access to the most recent ABAP Cloud Language Version 5. The image is used unchanged to ensure a consistent and comparable test basis. The benchmark process itself is implemented in Python. Communication between the Python control and the ABAP environment takes place via a python implementation of the \emph{ADT} interface, which enables programmatic access to the development environment. Communication with the various LLMs takes place via the OpenAI Python library, which enables connection to the respective models via various providers. An exception is Claude, for which the Claude Python library is required to use the Batch API.

Within a single test run, the Python script first sends a request to the respective LLM being investigated to generate ABAP source code for a defined programming task. A new ABAP class is then automatically created on the SAP server, the name of which is derived from the model's answer. The generated code is inserted into this class, saved, and activated. The system then creates a predefined ABAP unit test class, which serves to check the functionality and correctness of the generated program code. After executing the tests on the server, the results are returned to the Python script, which analyzes them, evaluates the success or failure of the test run, and if necessary identifies the underlying error type.

Errors can occur in various phases, for example during class creation, activation, during the syntax check, when inserting the source code, or during test execution. All error types are systematically logged and integrated into the subsequent evaluation. If an error occurs, the LLM is queried again to generate a corrected version of the program using the error message from the SAP server. This query process can include up to five iterations or ends prematurely as soon as an error-free solution is reached. This is determined by the SAP system through its syntax verification and our unit tests for semantic verification. Finally, the generated code is removed from the development environment to avoid influencing subsequent test cases.

In practical implementation, this process takes place in the form of batch processes. First, all requests are sent to the LLMs and the answers are saved locally before execution on the SAP server takes place and the results are returned to Python. Then the next batch with LLM requests, execution, and evaluation can be carried out. This division of the procedure enables a clear separation between the generation and execution of the test cases, which brings advantages in execution time and stability.

\subsection{Evaluation Criteria}
\label{sec:evaluationskriterien}
The evaluation of the generated solutions is based on a systematic investigation of success and failure. The primary measurement criterion is functional correctness, which is verified by passing automated ABAP unit tests on the SAP server.

To enable a differentiated assessment, the results are broken down according to the following aspects:

\begin{itemize}
    \item Error Classification: In the event of a failure, the reason for the failure is identified. A distinction is made as to which phase the error occurred in: during class creation, during the syntax check, or only during the execution of the unit tests. In addition, a detailed analysis of syntax errors is carried out to create characteristic error profiles of the models (e.g. declaration or structural errors).
    \item Iterative Success: It is measured after how many feedback iterations a model delivers a correct solution. This allows conclusions to be drawn about the learning curve and the effectiveness of compiler feedback for error correction and unit test success for semantic correctness.
    \item Model Comparison: The performance data is compared across all examined LLMs to show differences between proprietary and open-source architectures as well as specialized code models.
    \item Task Categories: The results are analyzed under the aspect of the five task categories introduced in Chapter \ref{sec:methodik_benchmark}. This approach makes it possible to evaluate the strengths and weaknesses of the models not only as a whole, but differentiated within specific problem types such as \textit{String Handling}, \textit{List or Array Operation}, \textit{Mathematical Calculation}, and \textit{Logical Condition}. The category of \textit{ABAP Database Operation} should be emphasized in particular, as these tasks come from a practical application context and are therefore of particularly high practical relevance for adapting business logic in SAP systems.
\end{itemize}

For the statistical evaluation of these time- or step-dependent processes, the Kaplan-Meier survival analysis is used \cite{kaplanNonparametricEstimationIncomplete1958, kleinbaumKaplanMeierSurvivalCurves2012}. Here, \say{survival} is defined as the persistence of the error state, i.e., the failure to pass the syntax check or the final unit tests. An \say{event} (in the sense of a \say{failure} or \say{death}) thus corresponds to the successful validation of the code. The analysis takes place in discrete time steps corresponding to the individual feedback iterations. In addition to estimating the survival probability, the method also provides 95\% confidence intervals, which allow a well-founded assessment of the significance and robustness of the performance differences.
The Kaplan-Meier method is particularly well suited for this type of analysis, as it models not only the temporal development of the probability of success over successive iterations, but also takes unsuccessful data into account \cite{kaplanNonparametricEstimationIncomplete1958, kleinbaumKaplanMeierSurvivalCurves2012}. The latter is relevant because not all test runs are successfully completed within the specified number of inquiries. With the help of the survival curves, it is thus possible to assess not only the final success but also the speed with which a model achieves a correct solution.

\section{Investigation}
\subsection{Results}
\subsubsection{Differences between Feedback Rounds}

\begin{figure}[h!]
    \centering
        \includegraphics[width=0.8\textwidth]{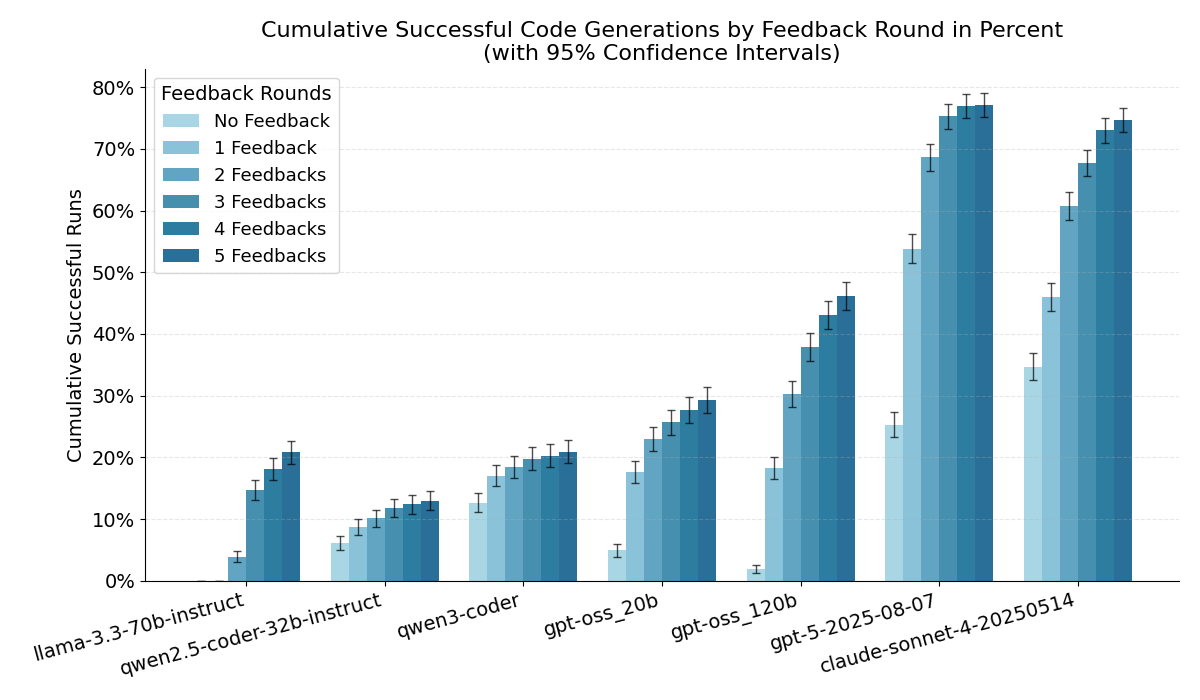}
        \caption{Benchmark results broken down by model and feedback rounds in percent.}
        \label{fig:success_by_feedbackRunde}
\end{figure}

\begin{figure}[h!]
    \centering
        \includegraphics[width=\textwidth]{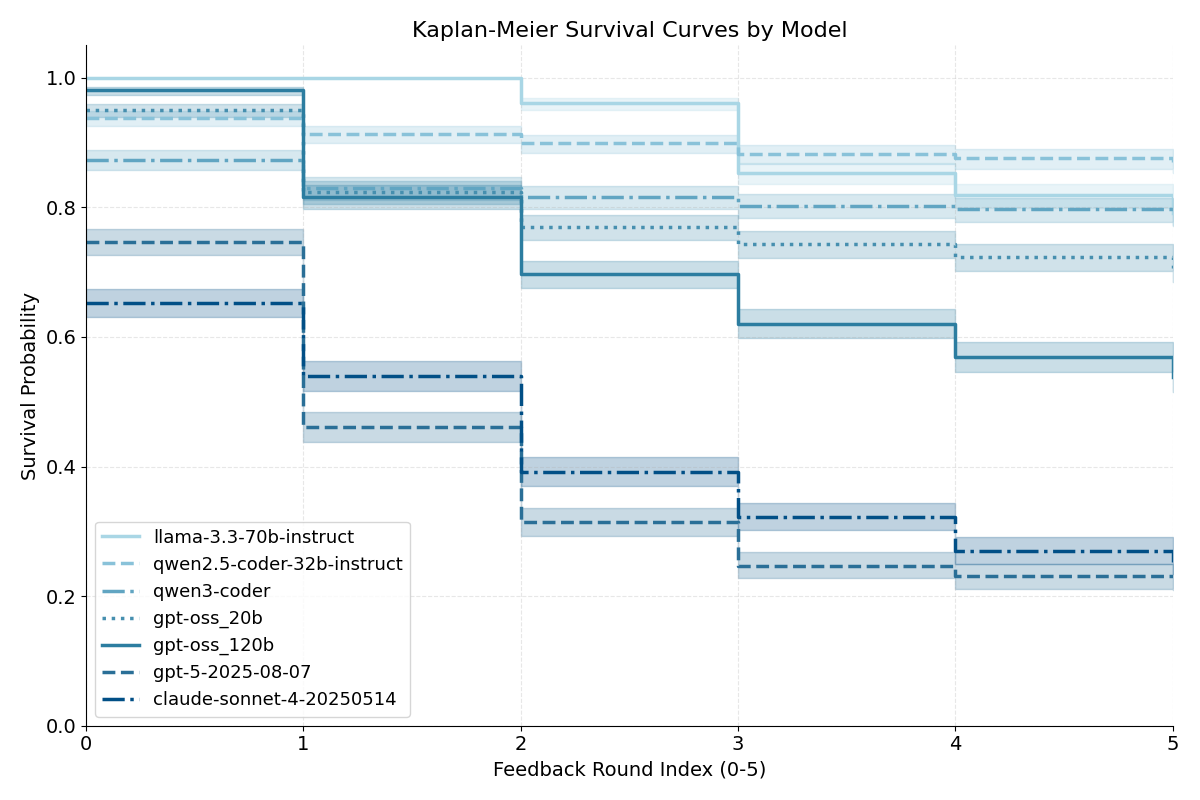}
        \caption{Kaplan-Meier survival curves showing the probability of remaining test errors over 0-5 feedback iterations (Basis: 10 x 180 test runs)}
        \label{fig:km_survival_curve_all_models}
\end{figure}

The results of Codestral-22B are not visualized in the figures \ref{fig:success_by_feedbackRunde}, \ref{fig:km_survival_curve_all_models} and \ref{fig:success_by_task_category} due to a success rate of 0\% across all feedback rounds, in order to ensure the readability. Figure \ref{fig:success_by_feedbackRunde} and Table \ref{tab:model_performance_by_feedback} provide an overview of the success rates of the other models across all feedback rounds. It is already visible here that the models react very differently to the feedback from the SAP compiler.

The analysis of the cumulative success rates across the feedback rounds illustrates the effectiveness of the iterative compiler feedback. Based on the Kaplan-Meier methodology described in Section \ref{sec:evaluationskriterien}, the survival curves in Figure \ref{fig:km_survival_curve_all_models} show the probability with which the models overcome the error state within the six observation intervals (rounds 0 to 5). The results reveal significant performance differences between the LLMs and show that all models benefit from the automated feedback, with the effectiveness of error correction correlating strongly with the exact model version.

GPT-5 and Claude-Sonnet-4 achieve by far the best results. While Claude-Sonnet-4 records the strongest entry with an average success rate of 24.1\% in the initial round (Round 0), GPT-5 demonstrates a superior learning curve in the subsequent feedback rounds and reaches the highest cumulative success rate of 77.1\% after the fifth feedback round. A significant proportion of successes is achieved for both models within the first two feedback rounds, with GPT-5 almost tripling its success rate from 19.28\% in Round 0 to 52.78\% in Round 2.

In contrast, open-source models such as GPT-OSS (120B and 20B) as well as the Qwen models show a lower basic competence in ABAP, but also benefit measurably from the feedback. Particularly noteworthy is the behavior of Llama-3.3-70B-Instruct, which does not deliver a single valid code contribution in the initial round, but achieves a cumulative success rate of 20.8\% through the feedback rounds. This underscores the relevance of automated feedback mechanisms for models with lower native domain expertise.

Figure \ref{fig:km_survival_curve_all_models} also shows that the greatest increase in success probability occurs in the first feedback iterations, especially for the more powerful models. At the same time, it can be seen that a measurable increase is still recorded in the fifth iteration. This suggests that with further feedback rounds, an additional increase in performance is possible and the learning curve has not yet completely flattened out.

\subsubsection{Differences between Task Types}

\begin{figure}[h!]
    \centering
    \includegraphics[width=\textwidth]{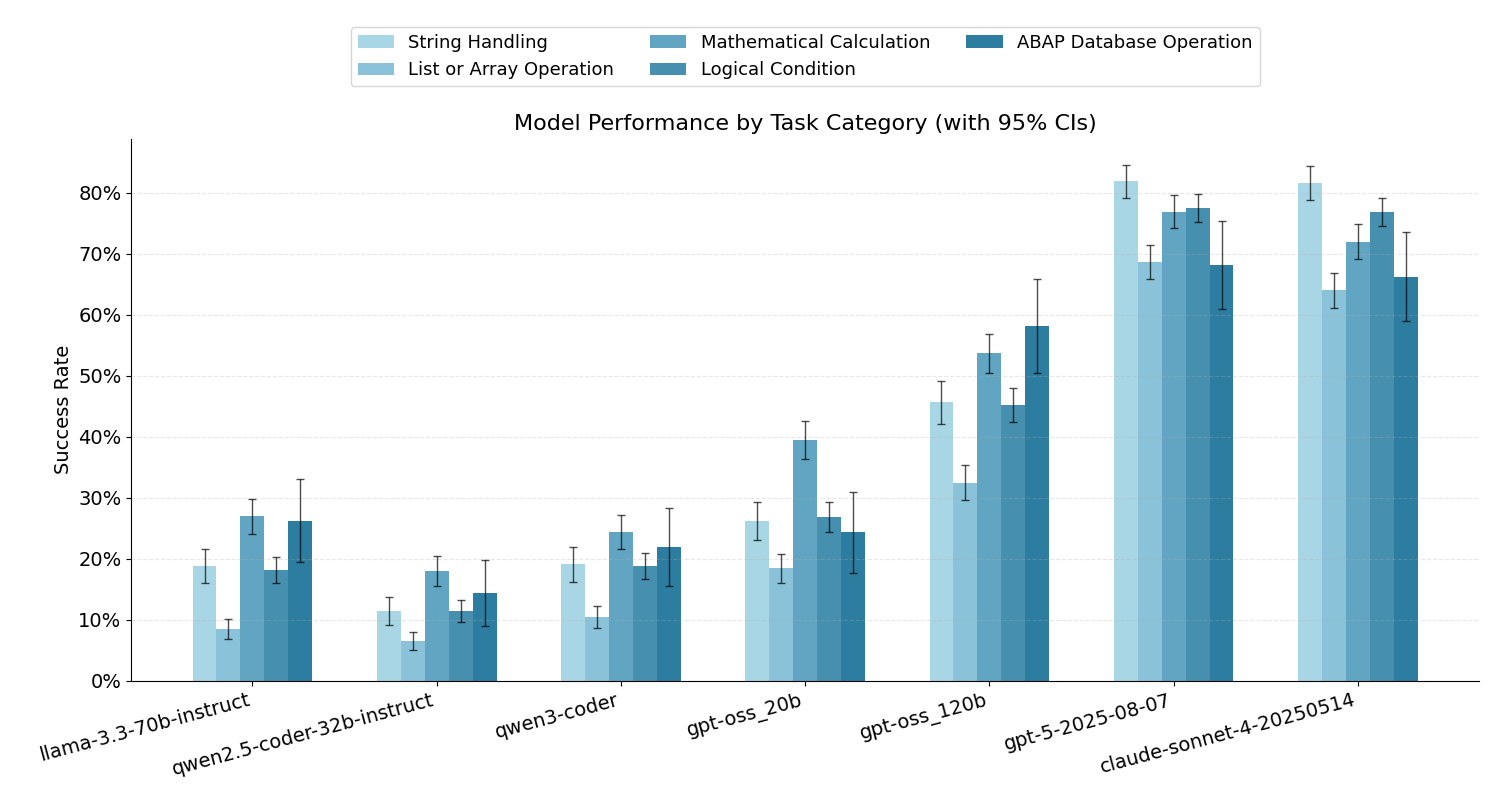}
    \caption{Benchmark results broken down by model and task focus in percent.}
    \label{fig:success_by_task_category}
\end{figure}

\begin{figure}[h!]
    \centering
        \includegraphics[width=0.7\textwidth]{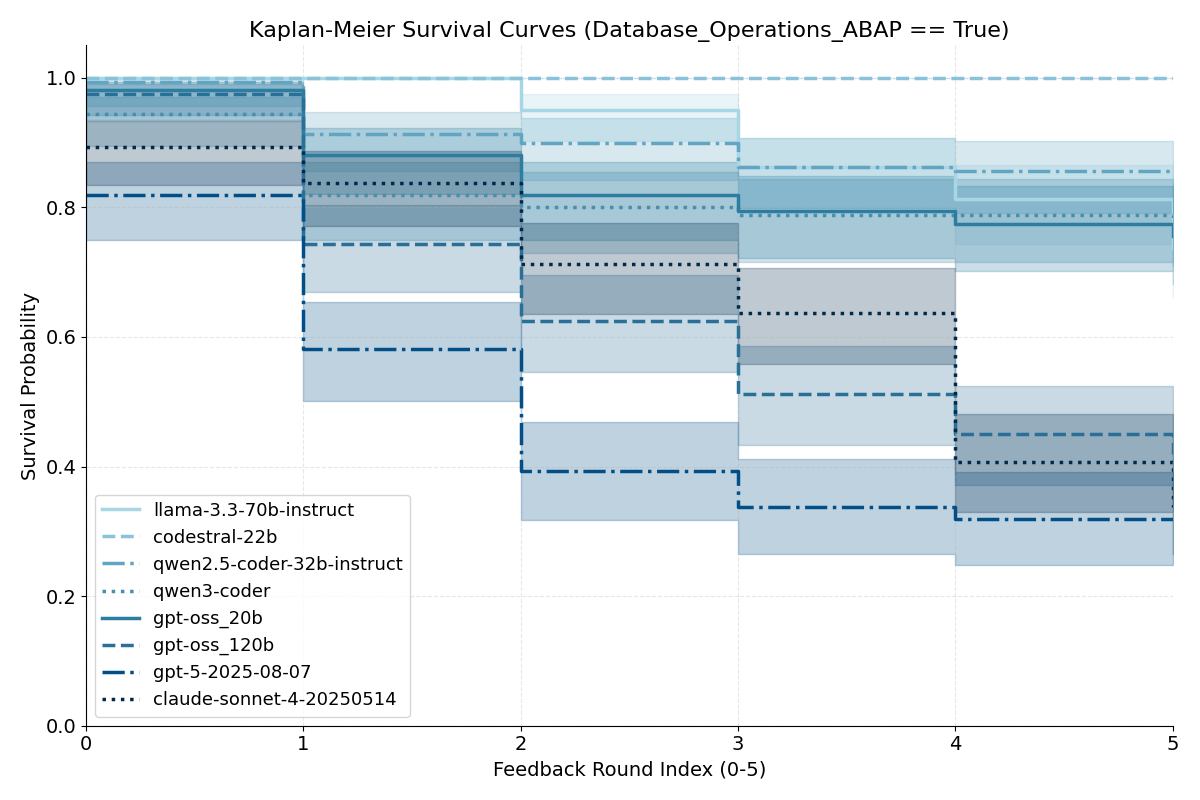}
        \caption{Kaplan-Meier survival curves for \textit{ABAP Database Operation} tasks differentiated by LLM models.}
        \label{fig:km_survival_curve_abap}
\end{figure}

Figure \ref{fig:success_by_task_category} and Table \ref{tab:model_performance_by_task_category} illustrate the success rates of the models differentiated by the five task categories: \textit{String Handling}, \textit{List or Array Operation}, \textit{Mathematical Calculation}, \textit{Logical Condition}, as well as \textit{ABAP Database Operation}. It turns out that performance remains consistent across most categories, with GPT-5 and Claude-Sonnet-4 in particular showing high robustness to different problem definitions. The smaller open-source models show increased susceptibility to errors in mathematical tasks and logical checks, while ABAP database procedures do not show an extraordinary increase in difficulty. The detailed model performance within the ABAP-specific subset is illustrated in Figure \ref{fig:km_survival_curve_abap}. This confirms the performance gap already observed.

Particularly striking is the behavior of GPT-5, which despite a moderate start in Round 0 shows highly efficient error post-processing and reaches a cumulative survival probability of approx. 0.32 (corresponds to approx. 68\% success). Claude-Sonnet-4 shows a somewhat lower initial performance in this subset with a starting value of approx. 89\% survival probability (11\% success) than in the overall average, but also approaches the top field through feedback. Also worth mentioning is the model Llama-3.3-70B-Instruct, which only achieves significant successes in this complex subcategory from the second feedback round onwards, which underlines the necessity of iterative feedback mechanisms for domain-specific tasks.

\begin{figure}[h!]
    \centering
        \includegraphics[width=0.6\textwidth]{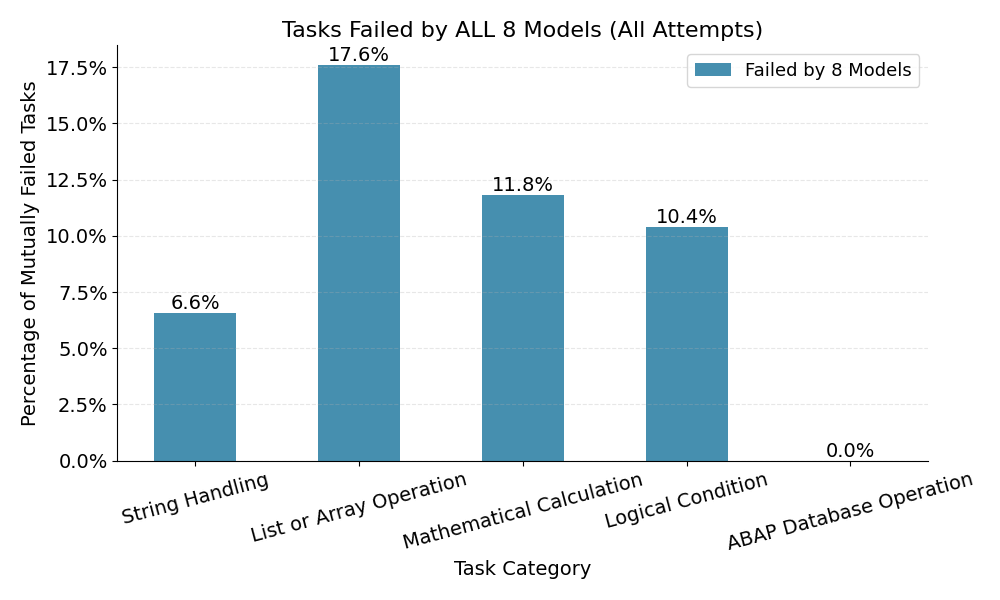}
        \caption{Percent of tasks per category that could not be solved by any of the eight models in all attempts.}
        \label{fig:failed_tasks_by_category}
\end{figure}

An analysis of tasks that none of the evaluated models could solve across all runs shows that the category \textit{List or Array Operation} poses the greatest challenge, with 19 unresolved cases out of 108. This is followed by \textit{Logical Condition} (13 out of 125) and \textit{Mathematical Calculation} (11 out of 93). Remarkably, in the category \textit{ABAP Database Operation}, every task was successfully solved by at least one model (see Figure \ref{fig:failed_tasks_by_category}). The full list of unsolved tasks is available in Appendix \ref{tab:failed_tasks}.

\begin{figure}[h!]
    \centering
        \includegraphics[width=0.8\textwidth]{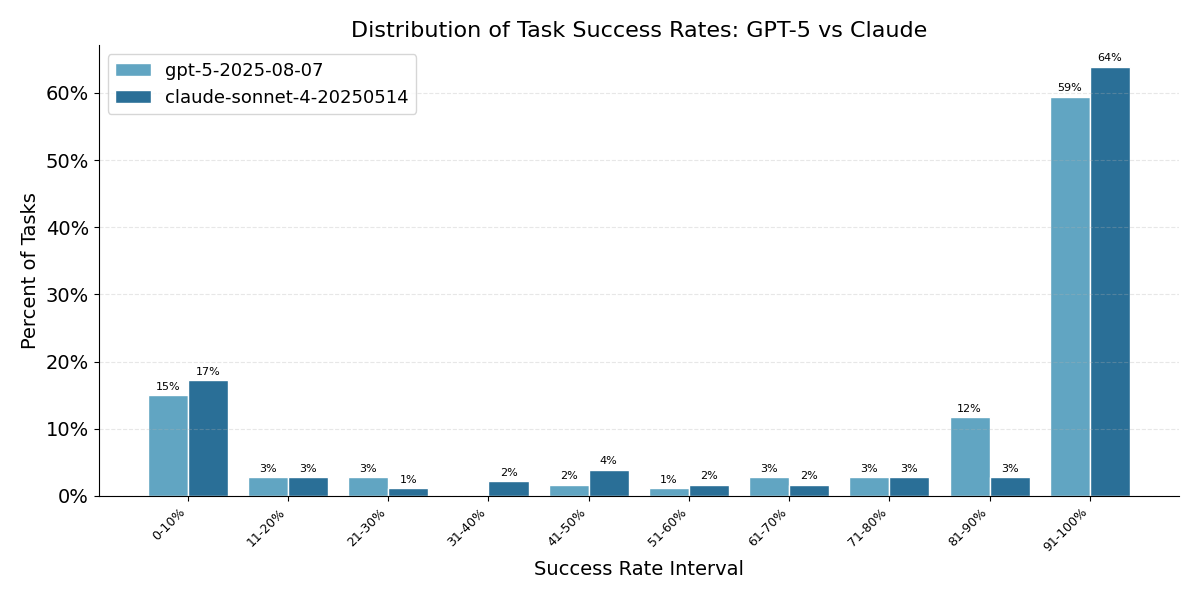}
        \caption{Frequency distribution of success rates per task for GPT-5 and Claude-Sonnet-4, illustrating the bimodal success characteristic.}
        \label{fig:success_distribution}
\end{figure}

The analysis of the success distribution per task for the two most powerful models, GPT-5 and Claude-Sonnet-4, also reveals a bimodal pattern (see Figure \ref{fig:success_distribution}). The models solve a task after five iterations either almost error-free (in 91-100\% of cases) or fail almost completely (in 0-10\% of cases). A \say{medium} success rate, where a task is solved in about half of the attempts, occurs only in exceptional cases. This speaks for a high determination of success by the specific task definition.

\subsubsection{Error Distribution}
\begin{figure}[h!]
     \centering
         \includegraphics[width=0.55\textwidth]{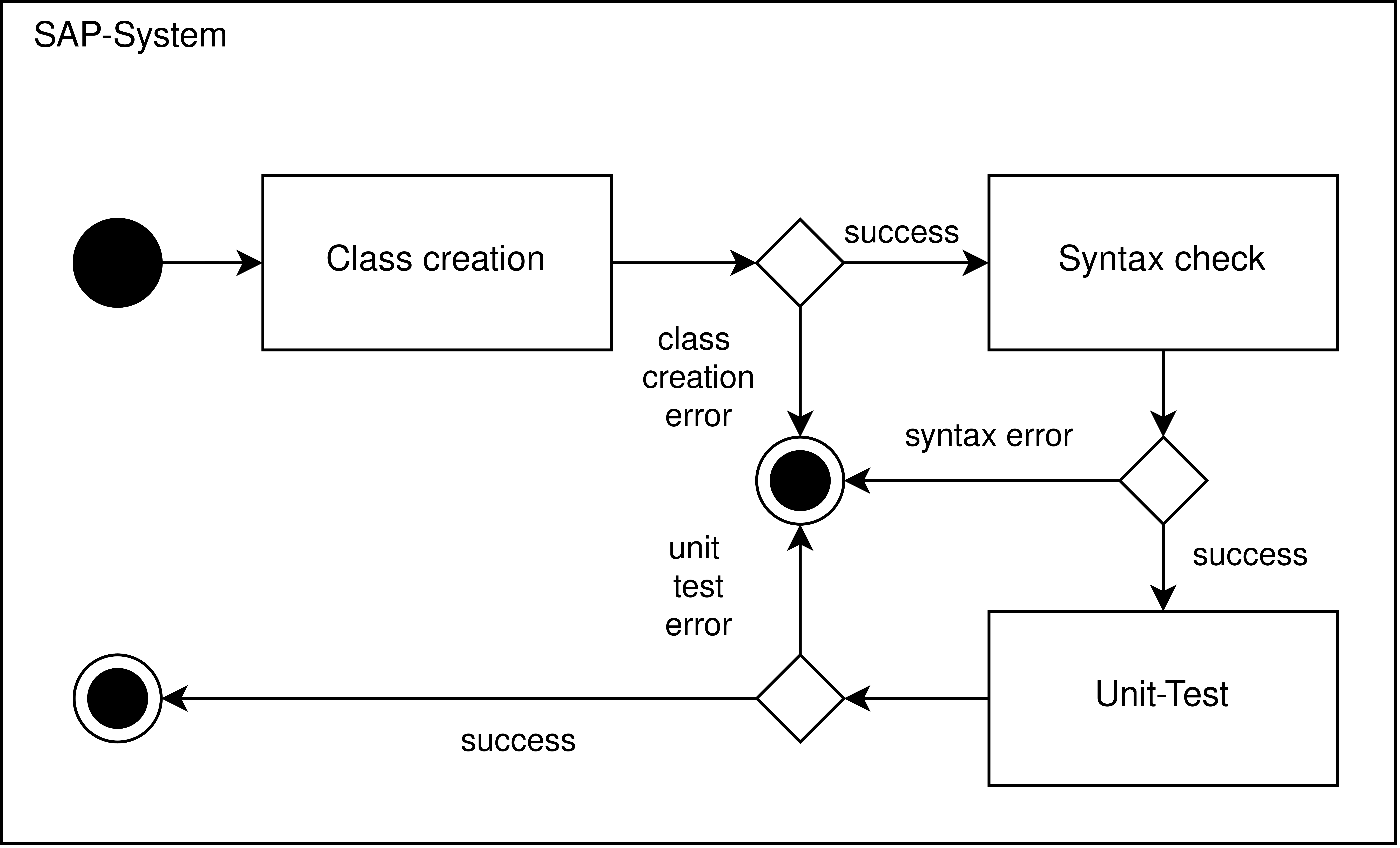}
         \caption{Flowchart of error checking in the SAP system.}
         \label{fig:fehlerablauf}
 \end{figure}

In the SAP system, errors are detected in a fixed order: First, class creation takes place, then a syntax check of the source code, and finally the execution of the unit tests, this is shown in Figure \ref{fig:fehlerablauf}. 

\begin{figure}[h!]
    \centering
        \includegraphics[width=0.7\textwidth]{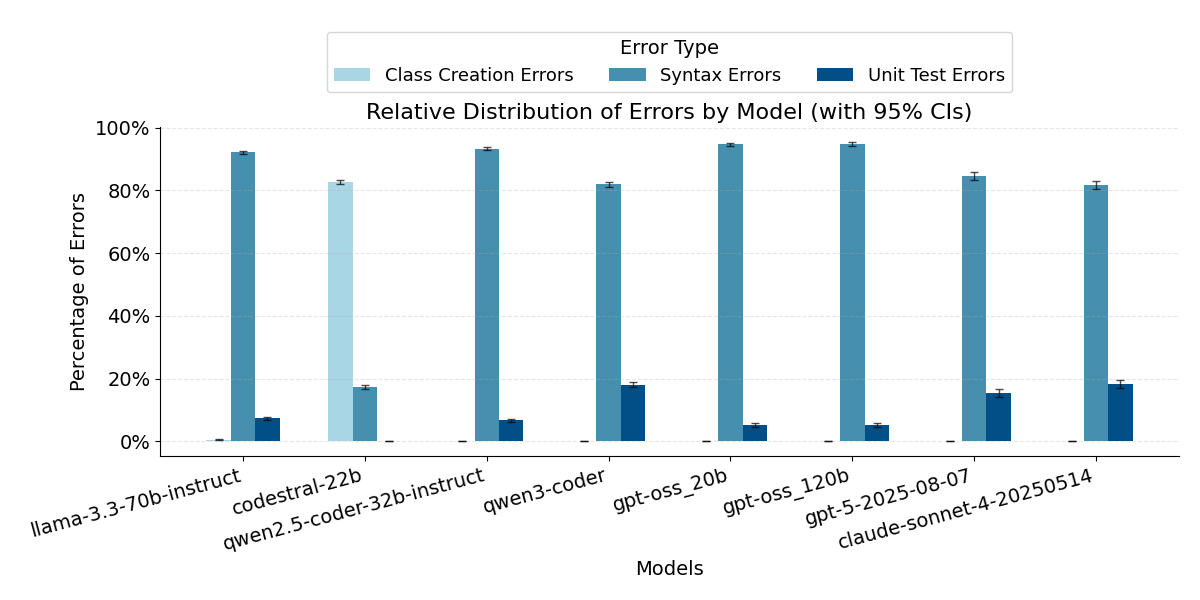}
        \caption{Distribution of error types by model in percent.}
        \label{fig:error_categories_by_model}
\end{figure}

Errors in early validation stages automatically prevent subsequent checks, so the interpretation of error rates depends on the respective validation stage. For error evaluation, it is only distinguished whether errors occur or not; especially with syntax errors, it is possible that several can occur, but here only the presence of the error category is counted. All values are shown in Table \ref{tab:model_error_types}. The analysis of the generated ABAP classes shows clearly different error patterns between the examined models (see Figure \ref{fig:error_categories_by_model}). 

Codestral-22B shows a particularly high error rate in class creation (82.72\%), while syntax errors (17.28\%) and unit test errors (0.00\%) never occur. This makes it clear that the model fails predominantly due to fundamental structural requirements. In contrast, GPT-5 generates ABAP classes with a low error rate in class creation (0.00\%) and a high syntax error rate (84.69\%). Consequently, only a smaller fraction of generations reaches the unit test phase, where an error rate of 15.31\% is observed. This indicates that GPT-5 is capable of producing structurally valid classes, but struggles primarily with syntactic correctness, while semantic issues in the unit tests play a secondary role. Other models show comparable error patterns, but differ in the validation stage at which they predominantly fail. GPT-OSS-120B and GPT-OSS-20B fail mainly at the syntax check, with error rates of 94.81\% and 94.78\% respectively, while class creation errors never occur in GPT-OSS-120B and and in 0.01\% of cases with GPT-OSS-20B. Llama-3.3-70B-Instruct and Qwen2.5-Coder-32B-Instruct reach the test phase more frequently, they show unit test errors in 7.40\% and 6.67\% of cases respectively. Qwen3-Coder and Claude-Sonnet-4 fail more often at the unit tests, with error rates of 18.01\% and 18.22\% respectively, indicating challenges in the semantic implementation of the required logic.

\begin{figure}[h!]
    \centering
        \includegraphics[width=0.9\textwidth]{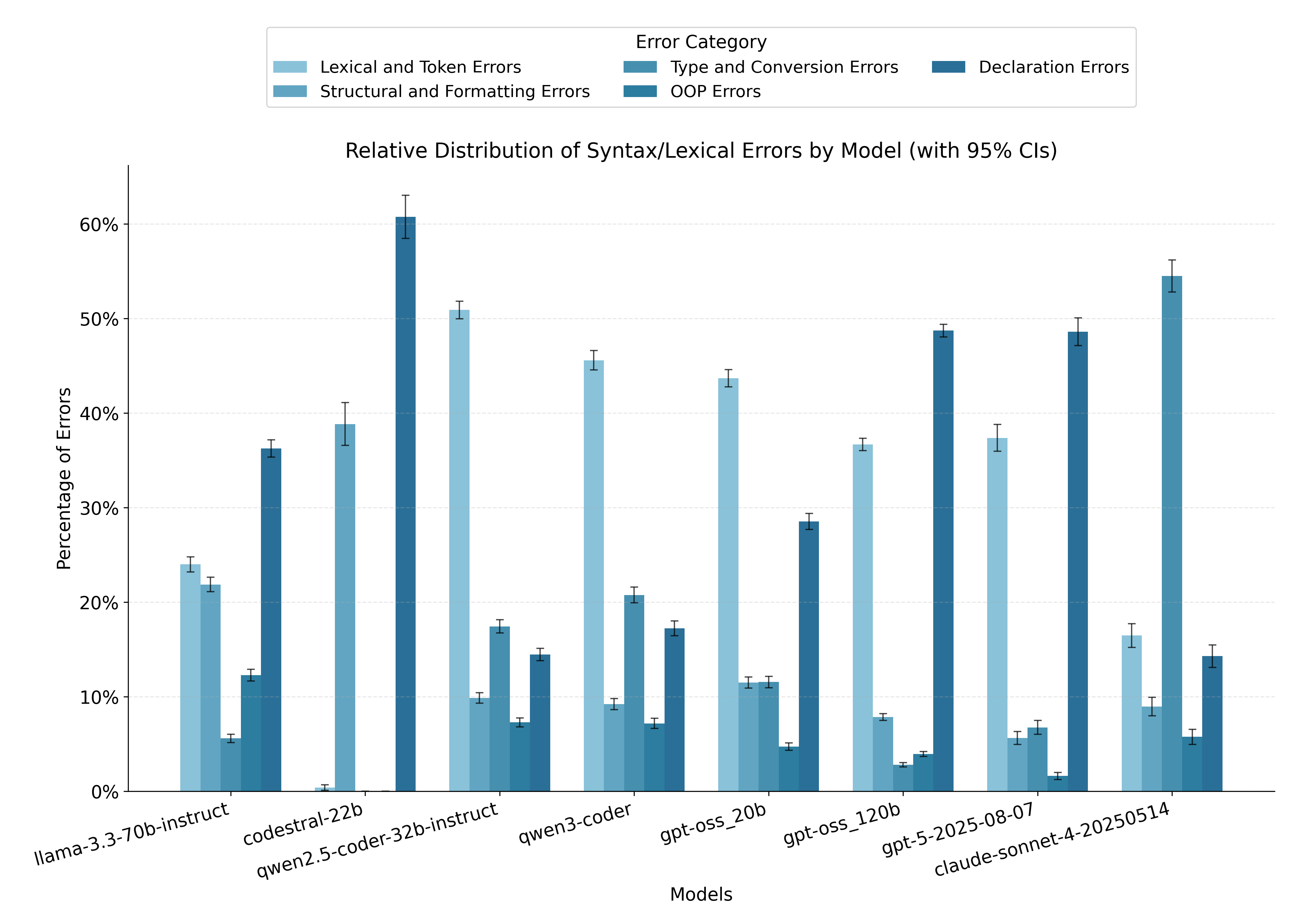}
        \caption{Distribution of syntax error categories by model in percent.}
        \label{fig:syntax_error_types_by_model}
\end{figure}

The detailed analysis of syntax errors, shown in Figure \ref{fig:syntax_error_types_by_model}, breaks down all syntax errors that occurred during the checks. The following types of errors are considered: 
\begin{itemize}
    \item Lexical and token errors refer to mistakes involving invalid characters or unexpected symbols in the code. 
    \item Structural and formatting errors involve issues with the overall code layout and organization, such as missing brackets or incorrect indentation. 
    \item Declaration errors occur when variables or data structures are improperly declared or omitted. 
    \item Type and conversion errors relate to mismatches or incorrect transformations between data types. 
    \item OOP-specific errors involve mistakes related to object-oriented programming concepts, such as class or method definitions. 
\end{itemize}

It is obvious that the models differ not only in the number but also in the types of errors generated. Llama-3.3-70B-Instruct has a comparatively balanced share of lexical and token errors (25.61\%) as well as structural and formatting errors (23.34\%). Qwen2.5-Coder-32B-Instruct, on the other hand, produces predominantly lexical and token errors (51.01\%), while structural errors account for only 9.91\%. Codestral-22B fails almost exclusively due to declaration errors (60.76\%) and structural problems (38.85\%); syntax errors themselves practically do not occur. GPT-5 shows moderate lexical errors (41.43\%) and a comparatively low error rate in OOP-specific aspects (1.79\%), while type and conversion errors (7.51\%) as well as declaration errors (43.02\%) dominate. Claude-Sonnet-4 generates predominantly type and conversion errors (55.17\%), while lexical and token errors accounted for only 16.62\%. The entire syntax error distribution is shown in Table \ref{tab:model_syntax_error_types}.

These results complement the validation stage analysis and clarify that the models not only fail with different frequencies at certain stages, but also exhibit characteristic error profiles in terms of syntax and structure.

\subsection{Evaluation}
The analysis carried out shows considerable potential for the use of LLMs for ABAP code generation. The results depend significantly on the ability to use compiler feedback for iterative improvement. Especially the more powerful models like GPT-5 and Claude-Sonnet-4 achieve success rates of over 74-77\% after five feedback cycles and demonstrate pronounced competence in dealing with error messages. The significant performance gains in early iterations, such as the increase of GPT-5 from 19.28\% to 58.67\% by the second round, illustrate the effectiveness of this process.

Another central finding is that the GPT-OSS models are also capable of achieving substantial improvements over several iterations. Despite lower model complexity, they manage to develop robust solutions with the help of feedback, which underscores the importance of the ability to adapt to errors beyond pure model size. The Qwen-Coder models start with comparatively good results in the first iteration, but show hardly any performance increases in the following rounds. This stagnation suggests that while they can safely reproduce basic ABAP structures, they are only limitedly capable of using the semantic information from compiler error messages for targeted optimizations. Llama-3.3-70B-Instruct represents a contrast to this: The model starts with weak performance, but improves significantly throughout the iterations and ultimately exceeds a success rate of at least 20.83\%, which indicates existing basic knowledge that can be applied to ABAP syntax through iterative feedback.
Special abnormalities are shown by the model Codestral-22B. It does not reliably adhere to the required Markdown code blocks and does not consistently separate explanatory text and code from each other. As a result, a clear extraction of the generated ABAP code is hardly possible, which significantly complicates validation and limits the usability of the answers. The difficulty lies therefore less in the code quality itself than in the lack of compliance with formal prompt specifications.

The analysis of the task types shows that their difficulty varies: While \textit{List or Array Operation} in particular poses structural challenges for the models (with 19 collectively unsolved tasks), the domain-specific \textit{ABAP Database Operation} are mastered by at least one model in every case. Nevertheless, the influence of the task type remains moderate compared to model quality. GPT-5 and Claude-Sonnet-4 achieve solid results even in complex categories, suggesting that model architecture and the quality of training data are the decisive factors for success. Why certain models perform significantly better in specific categories than others remains the subject of further investigations, where in particular the depth of domain knowledge in the training data could play a role.

Overall, the investigation shows that the successful use of LLMs in ABAP projects largely depends on a model's ability to use compiler feedback effectively. Some models deliver solid results initially but hardly improve, while others go through clear learning curves despite a weak start. At the same time, the results indicate that there are certain tasks for which even repeated compiler feedback does not lead to meaningful improvements, suggesting inherent limitations in the models ability to resolve specific problem types.

\section{Conclusion and Outlook}
\subsection{Conclusion}

The findings of our study show that benchmarking is an appropriate appproach to measure the capabilities of LLM performance for ABAP code generation. Although the field of considered LLMS and the amount of ABAP benchmark tasks are very limited the results suggest that LLMs are currently only capable to a limited extent of generating syntactically and semantically correct ABAP code without iterations. The best models, GPT-5 and Claude-Sonnet-4, achieved 19.28\% and 24.11\% correct answers to the first request. The other models remained well below this and partly delivered hardly any or no usable results. Thus, the first research question can be answered to the effect that LLMs can basically generate ABAP code, but the achievable quality without iteration is strongly model-dependent and for all systems lies far below a level relevant for practical use.

Regarding the second research question, iterative compiler feedback proves to be a decisive lever for quality improvement, with effectiveness depending significantly on the models. 
While the frontrunners GPT-5 and Claude-Sonnet-4 achieve a performance level after five iterations that brings support in interactive development processes within reach, the other models show significantly larger performance gaps. Despite measurable progress, Llama-3.3-70B-Instruct and the Qwen models achieve only low success rates, which strongly limits their practical usability. Particularly problematic is the performance of Codestral-22B, which failed completely across all rounds due to formal deficits in following prompt specifications.
For practical use, this means that currently above all models like GPT-5 and Claude-Sonnet-4 in combination with iterative feedback can offer substantial support to an SAP developer. 

The third research question dealt with the influence of the task type on the difficulty of ABAP code generation. The analysis shows that certain task categories pose significantly greater challenges for LLMs than others. In particular, \textit{List or Array Operation} proved to be problematic and, with 19 collectively unsolved tasks, had the highest error rates, indicating difficulties in correctly mapping complex control and data structures. In contrast, domain-specific \textit{ABAP Database Operation} could be successfully solved by at least one model in all cases. Overall, however, it turns out that the influence of the task type is moderate compared to the performance differences between the models. High-performance models like GPT-5 and Claude-Sonnet-4 achieve stable results even in complex categories, indicating that model architecture and training data quality represent the decisive factors for the success of code generation. Even if the task type thus has a measurable influence on the success rates, this is less significant compared to the very clear model differences. The choice of LLM therefore remains the decisive factor for the achievable code quality.

Despite the statistically significant improvements through iterations, the investigation reveals clear limits: A bimodal success pattern suggests that feedback primarily overcomes syntactic hurdles. However, if the model lacks the fundamental logical understanding for a task, even a high number of iterations hardly leads to a solution. With 19 collectively unsolved tasks, a significant core of algorithmic problems remains, which were not mastered for any of the examined LLMs. Consequently, this means that while LLMs have high potential for assisted workflows, they are currently not reliable enough for fully autonomous code generation in the ABAP environment.

\subsection{Outlook}
The present investigation has several limitations that must be taken into account when classifying the results. Since this is a still sparsely researched field, there are so far few scientific works on ABAP code generation with LLMs, which complicates both the theoretical embedding and the comparison with existing results. In addition, established, ABAP-specific benchmarks for standardized performance evaluation are missing. The empirical scope of the investigation is also limited, especially with regard to tasks that require direct integration of real SAP systems and database tables. A more comprehensive evaluation could provide additional insights here, especially the limited task amount and variation focused on ABAP-specific tasks should be further improved for future research.

The results on the third research question show that certain task types, especially those with complex data and control structures, pose special challenges for LLMs. However, the causes for this could not be conclusively clarified within the scope of this work. Future research should therefore investigate which linguistic or conceptual properties of these tasks lead to the observed difficulties and what role domain-specific knowledge plays in the training data. In addition, the influence of temperature as a generation parameter represents a relevant starting point for further investigations. In this work, a fixed temperature was used, so that possible effects on the success distribution remain unconsidered. It is to be expected that higher temperatures lead to a stronger variance of the results and this is reflected in an approximation to the middle of the success distribution shown in \ref{fig:success_distribution}.

At the same time, these limitations illustrate the considerable development potential of the research field. The continuous further development of LLMs as well as the increasing integration of AI into the SAP ecosystem, for example through tools like Joule for Developers, suggest that the quality of ABAP code generation will continue to increase. In particular, the direct connection to SAP systems and access to high-quality ABAP data offer great potential for this. In perspective, extensive automation of iterative code improvement processes also appears realistic.

From these considerations, several further research questions arise, including the suitability of further LLMs for ABAP code generation, the targeted handling of task-specific weaknesses, and the practical implementation of fully automated, feedback-based code generation and validation processes.

\section{Acknowledgments}
The authors gratefully acknowledge the computing time granted by the \say{KI-Servicezentrum für Sensible und Kritische Infrastrukturen} (KISSKI) and by the \say{THK-AI Forschungscluster} as well as organizational support by Stefan Breuer and Pascal Cerfontaine.

\clearpage

\bibliographystyle{IEEEtran}
\bibliography{main.bib}

\newpage
\appendix

\section{Appendix}

\label{github}
The code used is available on GitHub:\\
\url{https://github.com/timkoehne/LLM-Benchmark-ABAP-Code-Generation}

\label{huggingface}
The dataset used is available on HuggingFace:\\
\url{https://huggingface.co/datasets/timkoehne/LLM-ABAP-Code-Generation-Benchmark}

\begin{listing}[h!]
\begin{lstlisting}
You are an expert ABAP developer specializing in generating high-quality ABAP classes for SAP systems.
You write global ABAP classes with static methods.

System Environment
SAP Version: NetWeaver 7.57 (S/4HANA 2022)
Decimal Separator: Always use . (period) as decimal separator in numeric literals.

Development Rules
- Always use RETURNING parameters instead of EXPORTING.
- There should only be one public method per class.
- Class names must always start with Z.

Always respond with both the class definition and implementation code.
Only respond with the code. Do not include any explanations or comments.
\end{lstlisting}
\caption{System prompt used for ABAP code generation with LLMs}
\label{lst:abap-systemprompt}
\end{listing}

\begin{table}[h!]
    \centering
    \resizebox{\textwidth}{!}{
    \begin{tabular}{lcccccc}
    \hline
    \textbf{Model} & \textbf{Round 0} & \textbf{Round 1} & \textbf{Round 2} & \textbf{Round 3} & \textbf{Round 4} & \textbf{Round 5} \\
    \hline
    Llama-3.3-70B-Instruct & 0.00\% & 0.00\% & 3.67\% & 12.17\% & 16.89\% & 20.83\% \\
    Codestral-22B & 0.00\% & 0.00\% & 0.00\% & 0.00\% & 0.00\% & 0.00\% \\
    Qwen2.5-Coder-32B-Instruct & 6.00\% & 8.50\% & 10.06\% & 11.72\% & 12.44\% & 13.00\% \\
    Qwen3-Coder & 11.72\% & 16.22\% & 17.89\% & 19.28\% & 20.17\% & 20.94\% \\
    GPT-OSS-20B & 3.11\% & 12.94\% & 20.06\% & 24.39\% & 26.78\% & 29.33\% \\
    GPT-OSS-120B & 1.44\% & 13.61\% & 26.33\% & 35.28\% & 41.78\% & 46.17\% \\
    GPT-5 (2025-08-07) & 19.28\% & 41.83\% & 58.67\% & 68.56\% & 74.22\% & 77.11\% \\
    Claude-Sonnet-4 (2025-05-14) & 24.11\% & 41.50\% & 52.78\% & 63.17\% & 69.83\% & 74.67\% \\
    
    \hline
    \end{tabular}
    }
    \caption{Performance progression of various models across feedback rounds.}
    \label{tab:model_performance_by_feedback}
\end{table}

\begin{table}[h!]
    \centering
    \resizebox{\textwidth}{!}{
    \begin{tabular}{lccccc}
    \hline
    \textbf{Model} &
    \textbf{String Handling} &
    \textbf{List or Array Operation} &
    \textbf{Mathematical Calculation} &
    \textbf{Logical Condition} &
    \textbf{ABAP Database Operation} \\ 
    \hline
    Llama-3.3-70B-Instruct & 18.82\% & 8.52\% & 26.99\% & 18.24\% & 26.25\% \\ 
    Codestral-22B & 0.00\% & 0.00\% & 0.00\% & 0.00\% & 0.00\% \\ 
    Qwen2.5-Coder-32B-Instruct & 11.45\% & 6.48\% & 17.96\% & 11.44\% & 14.37\% \\ 
    Qwen3-Coder & 19.08\% & 10.46\% & 24.41\% & 18.80\% & 21.88\% \\ 
    GPT-OSS-20B & 26.18\% & 18.43\% & 39.46\% & 26.80\% & 24.37\% \\ 
    GPT-OSS-120B & 45.66\% & 32.50\% & 53.66\% & 45.20\% & 58.13\% \\
    GPT-5 (2025-08-07) & 81.84\% & 68.61\% & 76.88\% & 77.52\% & 68.13\% \\
    Claude-Sonnet-4 (2025-05-14) & 81.58\% & 63.98\% & 71.94\% & 76.80\% & 66.25\% \\ 
    \hline
    \end{tabular}
    }
    \caption{Performance metrics of various models across different coding task categories.}
    \label{tab:model_performance_by_task_category}
\end{table}

\begin{table}[h!]
    \centering
    \resizebox{\textwidth}{!}{
    \begin{tabular}{lrrr}
    \hline
    \textbf{Model} & \textbf{Class Creation Errors} & \textbf{Syntax Errors} & \textbf{Unit Test Errors} \\
    \hline
    Llama-3.3-70B-Instruct & 0.47\% & 92.13\% & 7.40\% \\
    Codestral-22B & 82.72\% & 17.28\% & 0.00\% \\
    Qwen2.5-Coder-32B-Instruct & 0.05\% & 93.29\% & 6.67\% \\
    Qwen3-Coder & 0.00\% & 81.99\% & 18.01\% \\
    GPT-OSS-20B & 0.01\% & 94.78\% & 5.20\% \\
    GPT-OSS-120B & 0.00\% & 94.81\% & 5.19\% \\
    GPT-5-2025-08-07 & 0.00\% & 84.69\% & 15.31\% \\
    Claude-Sonnet-4-20250514 & 0.00\% & 81.78\% & 18.22\% \\
    \hline
    \end{tabular}
    }
    \caption{Response errors by type of various models.}
    \label{tab:model_error_types}
\end{table}

\begin{table}[h]
    \centering
    \begin{tabular}{|c|l|}
    \hline
    Prompt ID & Category \\
    \hline
    0 & List or Array Operation, Mathematical Calculation, Logical Condition \\
    4 & List or Array Operation, Mathematical Calculation \\
    6 & String Handling, List or Array Operation, Mathematical Calculation, Logical Condition \\
    8 & List or Array Operation, Mathematical Calculation \\
    20 & List or Array Operation, Mathematical Calculation, Logical Condition \\
    21 & List or Array Operation, Mathematical Calculation \\
    22 & List or Array Operation, Logical Condition \\
    26 & List or Array Operation, Logical Condition \\
    28 & String Handling, List or Array Operation \\
    34 & List or Array Operation \\
    35 & List or Array Operation, Mathematical Calculation, Logical Condition \\
    61 & String Handling, Logical Condition \\
    81 & List or Array Operation, Logical Condition \\
    85 & List or Array Operation, Mathematical Calculation, Logical Condition \\
    88 & List or Array Operation, Mathematical Calculation, Logical Condition \\
    90 & List or Array Operation \\
    104 & String Handling, List or Array Operation, Logical Condition \\
    129 & List or Array Operation, Mathematical Calculation, Logical Condition \\
    135 & List or Array Operation, Logical Condition \\
    145 & String Handling, List or Array Operation, Mathematical Calculation \\
    \hline
    \end{tabular}
    \caption{List of Prompt IDs that failed across all models. Prompt content available at \ref{huggingface}.}
    \label{tab:failed_tasks}
\end{table}

\begin{table}[h!]
    \centering
    \resizebox{\textwidth}{!}{
    \begin{tabular}{lrrrrr}
    \hline
    \textbf{Model} & \textbf{Lexical \& Token Errors} & \textbf{Structural \& Formatting Errors} & \textbf{Type \& Conversion Errors} & \textbf{OOP Errors} & \textbf{Declaration Errors} \\
    \hline
    Llama-3.3-70B-Instruct & 25.61\% & 23.34\% & 6.61\% & 13.10\% & 31.34\% \\
    Codestral-22B & 0.40\% & 38.85\% & 0.00\% & 0.00\% & 60.76\% \\
    Qwen2.5-Coder-32B-Instruct & 51.01\% & 9.91\% & 17.77\% & 7.35\% & 13.97\% \\
    Qwen3-Coder & 46.00\% & 9.31\% & 23.34\% & 7.24\% & 14.11\% \\
    GPT-OSS-20B & 48.11\% & 12.67\% & 13.21\% & 5.30\% & 20.71\% \\
    GPT-OSS-120B & 39.33\% & 8.44\% & 3.07\% & 4.46\% & 44.70\% \\
    GPT-5-2025-08-07 & 41.43\% & 6.25\% & 7.51\% & 1.79\% & 43.02\% \\
    Claude-Sonnet-4-20250514 & 16.62\% & 9.07\% & 55.17\% & 5.87\% & 13.27\% \\
    \hline
    \end{tabular}
    }
    \caption{Syntax errors categorized for various models.}
    \label{tab:model_syntax_error_types}
\end{table}

\FloatBarrier
\setlength{\tabcolsep}{3pt} 
\begin{longtable}{
    |>{\centering\arraybackslash}p{0.09\textwidth}
    |>{\centering\arraybackslash}p{0.16\textwidth}
    |>{\centering\arraybackslash}p{0.18\textwidth}
    |>{\centering\arraybackslash}p{0.19\textwidth}
    |>{\centering\arraybackslash}p{0.18\textwidth}
    |>{\centering\arraybackslash}p{0.20\textwidth}|
}
\hline
\textbf{Nr} &
\textbf{String Handling} &
\textbf{List or Array Operation} &
\textbf{Mathematical Calculation} &
\textbf{Logical Condition} &
\textbf{ABAP Database Operation} \\
\hline
\endfirsthead

\hline
\textbf{Nr} &
\textbf{String Handling} &
\textbf{List or Array Operation} &
\textbf{Mathematical Calculation} &
\textbf{Logical Condition} &
\textbf{ABAP Database Operation} \\
\hline
\endhead

\hline
\caption{List of prompts/tasks and their category assignments.} \\
\endfoot

\hline
\caption{List of prompts/tasks and their category assignments.} \\
\endlastfoot

0 &  & X & X & X &  \\
1 & X & X & X & X &  \\
2 &  &  & X &  &  \\
3 &  & X & X & X &  \\
4 &  & X & X &  &  \\
5 &  & X &  & X &  \\
6 & X & X & X & X &  \\
7 & X & X &  & X &  \\
8 &  & X & X &  &  \\
9 &  & X & X & X &  \\
10 & X &  &  & X &  \\
11 & X & X &  & X &  \\
12 & X & X &  & X &  \\
13 &  &  & X & X &  \\
14 & X & X &  &  &  \\
15 & X & X &  &  &  \\
16 & X &  &  &  &  \\
17 & X & X &  & X &  \\
18 & X &  &  &  &  \\
19 & X & X &  & X &  \\
20 &  & X & X & X &  \\
21 &  & X & X &  &  \\
22 &  & X &  & X &  \\
23 & X &  &  &  &  \\
24 &  &  & X & X &  \\
25 &  & X & X & X &  \\
26 &  & X &  & X &  \\
27 & X &  &  &  &  \\
28 & X & X &  &  &  \\
29 & X & X &  & X &  \\
30 &  & X &  & X &  \\
31 &  &  & X & X &  \\
32 &  & X & X & X &  \\
33 &  & X & X & X &  \\
34 &  & X &  &  &  \\
35 &  & X & X & X &  \\
36 & X & X & X & X &  \\
37 &  & X &  & X &  \\
38 & X & X & X & X &  \\
39 &  & X & X & X &  \\
40 &  & X & X & X &  \\
41 &  &  & X &  &  \\
42 &  & X & X &  &  \\
43 &  & X & X & X &  \\
44 & X &  & X & X &  \\
45 &  &  & X &  &  \\
46 &  & X & X & X &  \\
47 &  & X & X & X &  \\
48 & X &  &  & X &  \\
49 &  &  & X &  &  \\
50 & X & X & X &  &  \\
51 & X &  &  &  &  \\
52 &  & X &  & X &  \\
53 &  &  & X &  &  \\
54 & X &  &  & X &  \\
55 &  &  & X &  &  \\
56 & X &  &  & X &  \\
57 &  & X &  & X &  \\
58 &  & X &  & X &  \\
59 &  &  & X & X &  \\
60 &  & X & X &  &  \\
61 & X &  &  & X &  \\
62 &  & X & X &  &  \\
63 &  &  & X &  &  \\
64 & X &  & X &  &  \\
65 & X &  & X &  &  \\
66 & X &  & X & X &  \\
67 & X &  & X &  &  \\
68 &  & X &  & X &  \\
69 &  & X &  & X &  \\
70 &  & X &  & X &  \\
71 &  &  & X & X &  \\
72 &  & X & X & X &  \\
73 &  & X &  & X &  \\
74 & X & X &  & X &  \\

75 &  &  & X & X &  \\
76 &  &  & X & X &  \\
77 &  &  & X & X &  \\
78 & X &  & X &  &  \\
79 & X &  & X &  &  \\
80 & X &  &  & X &  \\
81 &  & X &  & X &  \\
82 & X &  & X & X &  \\
83 &  &  & X &  &  \\
84 & X &  & X &  &  \\
85 &  & X & X & X &  \\
86 & X &  &  &  &  \\
87 &  & X &  & X &  \\
88 &  & X & X & X &  \\
89 & X &  &  &  &  \\
90 &  & X &  &  &  \\
91 & X &  &  & X &  \\
92 &  &  & X & X &  \\
93 & X &  &  & X &  \\
94 &  & X & X &  &  \\
95 & X &  &  & X &  \\
96 &  & X & X & X &  \\
97 &  &  & X &  &  \\
98 & X &  &  & X &  \\
99 & X &  &  & X &  \\
100 &  & X & X &  &  \\
101 & X & X &  & X &  \\
102 &  &  &  & X &  \\
103 & X &  & X &  &  \\
104 & X & X &  & X &  \\
105 & X & X &  & X &  \\
106 &  & X & X & X &  \\
107 & X &  &  & X &  \\
108 & X & X &  & X &  \\
109 &  & X &  & X &  \\
110 &  & X &  & X &  \\
111 & X &  &  &  &  \\
112 & X &  &  & X &  \\
113 & X & X &  &  &  \\
114 &  & X & X & X &  \\
115 &  & X &  & X &  \\
116 & X & X & X & X &  \\
117 & X & X &  & X &  \\
118 & X &  &  & X &  \\
119 & X &  &  & X &  \\
120 &  & X &  &  &  \\
121 &  & X & X & X &  \\
122 & X & X & X & X &  \\
123 &  & X & X & X &  \\
124 & X &  &  & X &  \\
125 & X & X &  & X &  \\
126 &  & X &  & X &  \\
127 &  &  & X & X &  \\
128 &  & X & X & X &  \\
129 &  & X & X & X &  \\
130 &  & X & X & X &  \\
131 & X &  & X & X &  \\
132 & X &  &  & X &  \\
133 &  & X & X &  &  \\
134 & X &  &  & X &  \\
135 &  & X &  & X &  \\
136 &  & X &  &  &  \\
137 & X &  &  & X &  \\
138 &  &  & X & X &  \\
139 &  &  & X &  &  \\
140 & X &  &  & X &  \\
141 & X &  &  & X &  \\
142 &  & X & X & X &  \\
143 & X &  & X & X &  \\
144 & X &  & X & X &  \\
145 & X & X & X &  &  \\
146 &  & X &  & X &  \\
147 &  & X & X & X &  \\
148 &  & X & X & X &  \\
149 & X & X &  &  &  \\
150 &  &  & X & X &  \\
151 &  & X & X & X &  \\
152 &  & X & X &  &  \\
153 & X & X &  & X &  \\
154 & X &  &  & X &  \\
155 & X & X &  & X &  \\
156 & X &  & X & X &  \\
157 &  &  & X & X &  \\
158 & X & X &  & X &  \\
159 &  & X & X & X &  \\
160 &  & X & X & X &  \\
161 & X &  &  &  &  \\
162 & X &  &  &  &  \\
163 &  & X &  & X &  \\

erp\_000 &  &  &  &  & X \\
erp\_001 &  & X &  &  & X \\
erp\_002 &  & X &  &  & X \\
erp\_003 &  & X &  &  & X \\
erp\_004 &  &  &  & X & X \\
erp\_005 &  & X & X & X & X \\
erp\_006 &  & X & X & X & X \\
erp\_007 &  & X & X & X & X \\
erp\_008 & X & X &  & X & X \\
erp\_009 &  & X & X & X & X \\
erp\_010 &  & X &  & X & X \\
erp\_011 &  & X & X & X & X \\
erp\_012 &  & X & X &  & X \\
erp\_013 &  & X &  & X & X \\
erp\_014 &  & X &  & X & X \\
erp\_015 &  & X & X &  & X \\
\hline
\end{longtable}

\end{document}